# Strain-Induced Non-alter Compensated Magnet and Its Application to Magnetic Tunnel Junction Device Design

Fangqi Liu[1#], Yanrong Song[1#], Zhenhua Zhang[2,3*], Yong Liu[1], Sicong Zhu[2,4,5], Zhihong Lu[2,3*]

and Rui Xiong[1*]

[1]Key Laboratory of Artificial Micro- and Nano-structures of Ministry of Education, School of Physics and Technology, Wuhan University, Wuhan 430072, People's Republic of China

[2]The State Key Laboratory of Refractories and Metallurgy, Wuhan University of Science and Technology, Wuhan 430081, People's Republic of China

[3]School of Material Science and Technology, Wuhan University of Science and Technology, Wuhan 430081, People's Republic of China

[4]Hubei Province Key Laboratory of Systems Science in Metallurgical Process, Wuhan University of Science and Technology, Wuhan 430081, People's Republic of China

[5]Department of Mechanical Engineering, National University of Singapore, Singapore 117575, Singapore

*Corresponding author: zzhua@wust.edu.cn

*Corresponding author: zludavid@live.com

*Corresponding author: xiongrui@whu.edu.cn

**ABSTRACTS**

The recent proposal of altermagnetism has drawn widespread attention to antiferromagnet (AFM) exhibiting spin splitting, extending beyond the realm of sign-alternating spin splitting in momentum space protected solely by rotational symmetry. Herrin, we propose a shear-strain strategy that enables significant modulation of d-wave altermagnets into an non-alter compensated magnets. A comprehensive analysis combining the magnetic moment compensation characteristics of opposite spin sublattices with the distribution of spin-resolved conduction channels in momentum space under the [001] crystal orientation reveals that shear strain breaks the rotational symmetry of alternatmagnets. To explore the application potential of non-alter compensated magnets, we design $RuO_2/TiO_2/RuO_2$ magnetic tunnel junctions (MTJ) with three crystallographic orientations ((001), (110), (100)) and investigated their transport properties under shear strain. This non-alter electronic structure not only enhances the tunneling magnetoresistance (TMR) in spin-split paths of intrinsic $RuO_2$ (226% to 431%) but also enables substantial TMR in spin-degenerate paths (from 0-88%)). Our work provides guidelines for broadening magnetic materials and device platforms.

**INTRODUCTION**

Antiferromagnetic (AFM) materials have unique advantages such as robustness to external magnetic fields, high-frequency dynamic characteristics on the picosecond time scale, and the disappearance of macroscopic net magnetization compared with ferromagnet (FM) materials[1-3]. However, due to the spin degeneracy in the electronic band structure of common AFM materials, tunneling magnetoresistance (TMR) in AFM tunnel junction (AFMTJ) based on these AFM is not feasible[4-8]. A novel magnetism—altermagnetism—transcending conventional FM and AFM orders has been proposed[9-12]. Altermagnets—characterized by non-relativistic alternating spin splitting in the band structure and collinear compensated magnetic moments—have recently garnered substantial attention for their fundamental physics and potential device applications[6, 13-15].

In conventional collinear AFM, opposing magnetic sublattices are related through the combined operation of time-reversal symmetry (*T*) and spatial symmetries—including space inversion (*P*) or half-unit-cell translation (*t*)—enforcing spin degeneracy across the entire Brillouin zone. In contrast, the spin motif of altermagnets connects counter-rotating sublattices via crystalline rotational symmetry (*R*). This rotational operation enforces sign-alternating spin splitting along momentum-space trajectories associated with R, resulting in band crossings at the

Γ-point. Consequently, spin-resolved bands intersect at the Brillouin zone center and remain degenerate at Γ[10, 16-18]. Altermagnets are classified by the higher-order rotational symmetry of spin sublattices into d-wave, g-wave, and i-wave types, leading to spin splitting exclusively along selected crystallographic directions[19-23]. For experimental realization of MTJ with large TMR, the transport axis must align with the spin-splitting directions of the altermagnet. This requirement imposes significant constraints on material growth and junction fabrication[24, 25].

The concept of spin-split AFM was first proposed by Pekar and Rashba in 1964. Subsequently, van Leuken et al. predicted in 1995 a half-metallic AFM exhibiting 100% spin polarization and spin splitting across high-symmetry paths traversing the entire Brillouin zone at the Fermi level—later experimentally observed[26-28]. However, such phenomena have been predominantly confined to Heusler alloys, limiting their practical deployment. Recent studies demonstrate that doping or electric-field gating in conventional AFM can engineer a compensated magnetic state with spin-split bands. Distinct from altermagnets, such modified systems exhibit uniform spin splitting throughout momentum space without sign alternation at the Γ-point[29, 30]. The fundamental mechanism involves external perturbations simultaneously breaking rotational symmetry between opposing spin sublattices and enforcing magnetization compensation.

In this work, we propose a shear-strain strategy to effectively break the rotational symmetry of d-wave altermagnets, transforming them into non-alter compensated magnets. Experimentally, such shear strain can be implemented via tilted epitaxial growth induced by controlled lattice mismatch. Both the magnitude and direction of the applied strain significantly enhance spin splitting and enable switchable magnetization. This approach dramatically boosts TMR in $RuO_2/TiO_2/RuO_2$ MTJ across arbitrary transport orientations.

**METHODS**

First-principles calculations of the electronic structure of the bulk electrode and the interlayer are carried out at the material level, while the transport properties of the MTJ are calculated at the device level. At the material level, first-principles calculations are performed based on density functional theory (DFT) as implemented in the Vienna Ab- initio Simulation Package (VASP) [31, 32]. At the device level, a two-terminal structure with $RuO_2/TiO_2/RuO_2$ sandwiched between two semi-infinite leads is employed to investigate the TMR effect. The transport performance of the MTJ is calculated using the DFT with nonequilibrium Green's function [33] formalism as implemented in QuantumATK [34]. The exchange-correlation potential is explained by the Perdew-Burke-Ernzerhof (PBE) functional based on spin-generalized gradient approximation plus on-site Coulomb interaction (SGGA+U) functional with $U_{eff}$ = 5 eV on Ti-3$d$, and $U_{eff}$ =2.5 eV for Ru-4$d$ [35, 36]. The projector augmented-wave method is used for wave function expansion with an energy cut-off of 450 eV. The geometry optimization continues until the energy differences and ionic forces converge to less than $10^{-6}$ eV and 0.01 eV/Å, respectively. Monkhorst-Pack k-point meshes of 12 × 12 × 12 are used for bulk $RuO_2$ and 11 ×11 × 100 for $RuO_2/TiO_2/RuO_2$ MTJ.

**Results**

In altermagnets, the breaking of spin-sublattice translational symmetry while preserving rotational symmetry leads to spin-resolved band crossings at the center of the Brillouin zone, resulting in spin degeneracy at the Γ point. Simultaneously, the specific pattern of spin splitting in an altermagnet is determined by the higher-order rotational symmetry (e.g., d-wave, g-wave) of the material type. Consequently, spin-degenerate paths coexist. Hypothetically, if the rotational symmetry between the two spin sublattices is further broken, it could lift the spin degeneracy not only at high-symmetry points but also at generic k-points (non-specific paths), while compensating the magnetization (maintaining net zero magnetization). Therefore, in this work, we employ a d-wave altermagnet as a model system and propose the application of shear strain to break the rotational symmetry of the counter-rotating spin sublattices in real space. Crucially, this biaxial shear strain does not disrupt the local atomic environment around the sublattices to the extent of inducing FM alignment (thus preserving the compensated antiparallel order).

Experimentally, tilted growth of materials can be achieved by leveraging the lattice mismatch between two materials, thereby introducing shear strain. For instance, epitaxial growth of Co on [111]-oriented Pt/SiO$_2$ substrate enables the tilted growth of Co[37]. Here, using RuO$_2$—a representative d-wave altermagnetic material—as an example, we apply biaxial shear strain ranging from -5% to +5%. Initial testing of RuO$_2$'s magnetic ground state reveals that it retains its AFM order under shear strain, consistent with our prior hypothesis: the applied shear strain does not significantly increase the disparity in the surrounding atomic environments between the opposing spin sublattices. Further investigation of its electronic structure reveals that spin splitting occurs throughout the entire Brillouin zone under shear strain. Significantly, this behavior is distinct from the characteristic spin-texture alternation around high-symmetry points observed in altermagnets. Further investigation of the strain dependence of the spin-splitting magnitude reveals distinct behaviors. At high-symmetry points (notably Γ), the spin splitting is initially zero in the unstrained state. As the shear strain increases, the spin-splitting energy increases linearly. The maximum spin-splitting energy reaches 0.65 eV at the X point. Remarkably, the magnetization along certain crystallographic directions in RuO$_2$ flips upon reversal of the applied strain direction, indicating strain-controlled magnetic switching.

Furthermore, we investigate the evolution of magnetic moments in RuO$_2$ under varying biaxial shear strain. In the absence of strain, RuO$_2$ exhibits a compensated magnetic state with zero net moment (note: spin-orbit coupling (SOC) is neglected here; altermagnets are reported to possess weak orbital moments). Upon application of shear strain, however, RuO$_2$ develops an extremely small net magnetic moment, ranging from -0.017 to 0.015 μB. The origin of this non-zero net moment is twofold: (1) the inequivalent magnetic moments of the anti-parallel Ru atoms, and (2) contributions from the oxygen atoms within the opposing sublattices. This observation directly demonstrates that the applied shear strain breaks the rotational symmetry between the sublattices, resulting in asymmetricelectron clouds around them. Moreover, this net magnetic moment is not the direct cause of the spin splitting observed in $RuO_2$, as its extremely small magnitude (on the order of 0.015 μB) cannot account for spin-splitting energies reaching hundreds of meV. We also applied biaxial shear strain to other d-wave altermagnets, such as $MnF_2$, and observed similar phenomena. This demonstrates the generality of our proposed strain-engineering strategy.

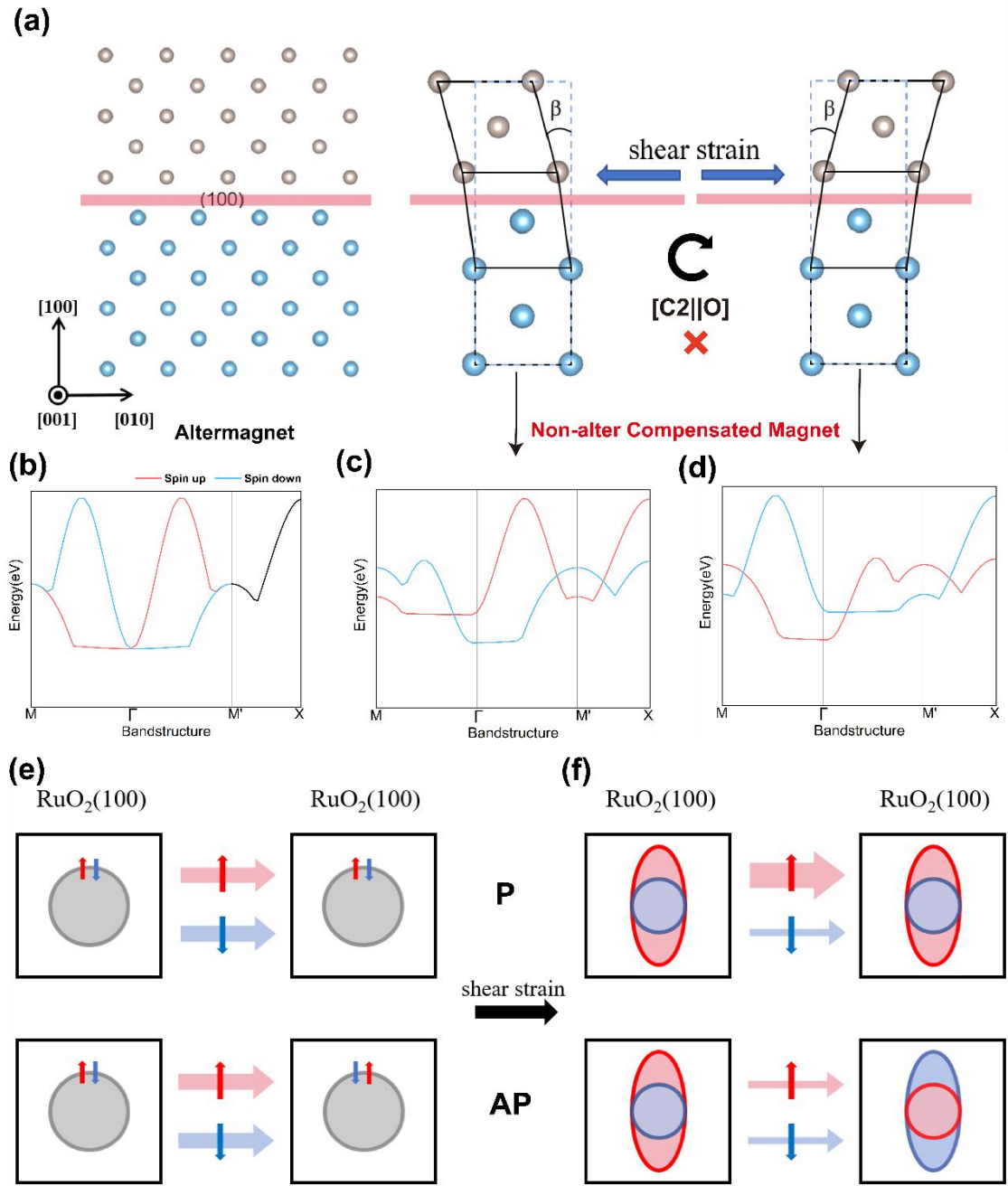

**Figure 1.** (a) The schematic diagram illustrates oblique-angle deposition for achieving shear strain,

where the upper and lower materials must maintain a lattice mismatch. Band structures of (b) altermagnets and (c)-(d) non-alter compensated magnets, with different shear strain directions enabling magnetization reversal in the non-alter compensated magnets. Spin-dependent transport properties in the RuO$_2$/TiO$_2$/RuO$_2$ MTJ along the [100] orientation (e) with and (f) without applied shear strain.

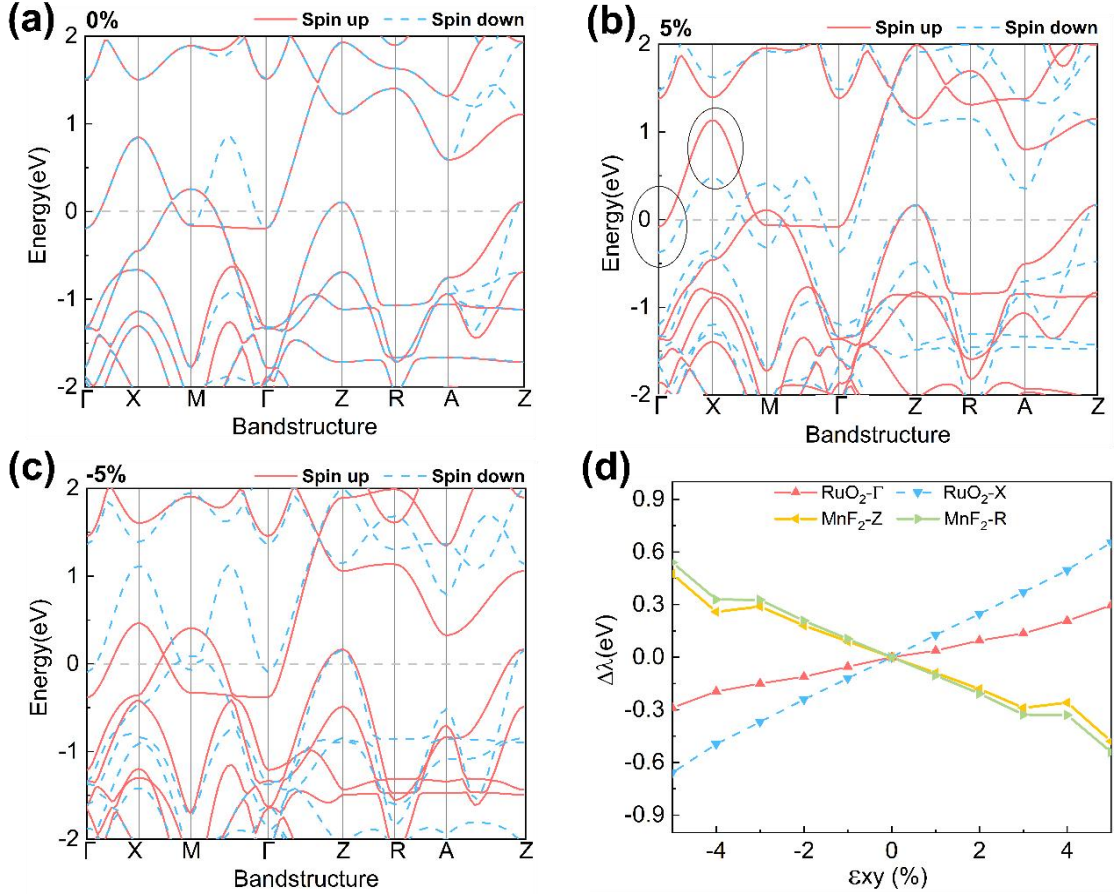

**Figure 2.** Band structures of (a) intrinsic RuO$_2$ with (b) +5% and (c) -5% shear strain. Dependence of spin-splitting strength on shear strain magnitude and direction at high-symmetry K-points in RuO$_2$ and MnF$_2$.

We next explore the device applicability of non-alter RuO$_2$. Significant advances have recently been achieved in AFMTJ research. The [001] crystallographic direction in RuO$_2$ constitutes a unique transport path. Band structure analysis along the Γ-Z path reveals spin degeneracy. However, momentum-space visualization demonstrates a pronounced mismatch between spin-up and spin-down transport channels. This mismatch is inherently compensated by rotational symmetry protection, resulting in overall spin-degenerate band manifestations (Figure 3(a)). Consequently, the

RuO$_2$/TiO$_2$/RuO$_2$ (001) MTJ exhibits substantial TMR, achieved by switching the matching conditions between spin-polarized transport channels in the two RuO$_2$ electrodes (Figure 3(c)). Subsequently, we investigated the spin-dependent electron transmission in momentum space of RuO$_2$ under applied shear strain. It can be found that as the shear strain increases, the difference in the number of conduction channels between spin-up and spin-down electrons becomes larger, and the rotational symmetry of these spin-resolved electron channels is significantly broken. Concurrently, as the direction of the strain is altered, the difference in the number of spin-up and spin-down conduction channels reverses, corresponding to the magnetization reversal observed in the preceding band structure analysis. However, the TMR of the shear-strain-based RuO$_2$/TiO$_2$/RuO$_2$ MTJ does not exhibit a significant increase with small shear strain (1%-3%). In contrast, under larger strains (4%-5%), the TMR increases markedly, rising from 226% to 431%. This behavior arises because the TMR along the [001] transport path primarily originates from the mismatch in spin transport channels. Furthermore, under small shear strain, the applied strain yields only a modest increase in the spin polarization along this [001] path.

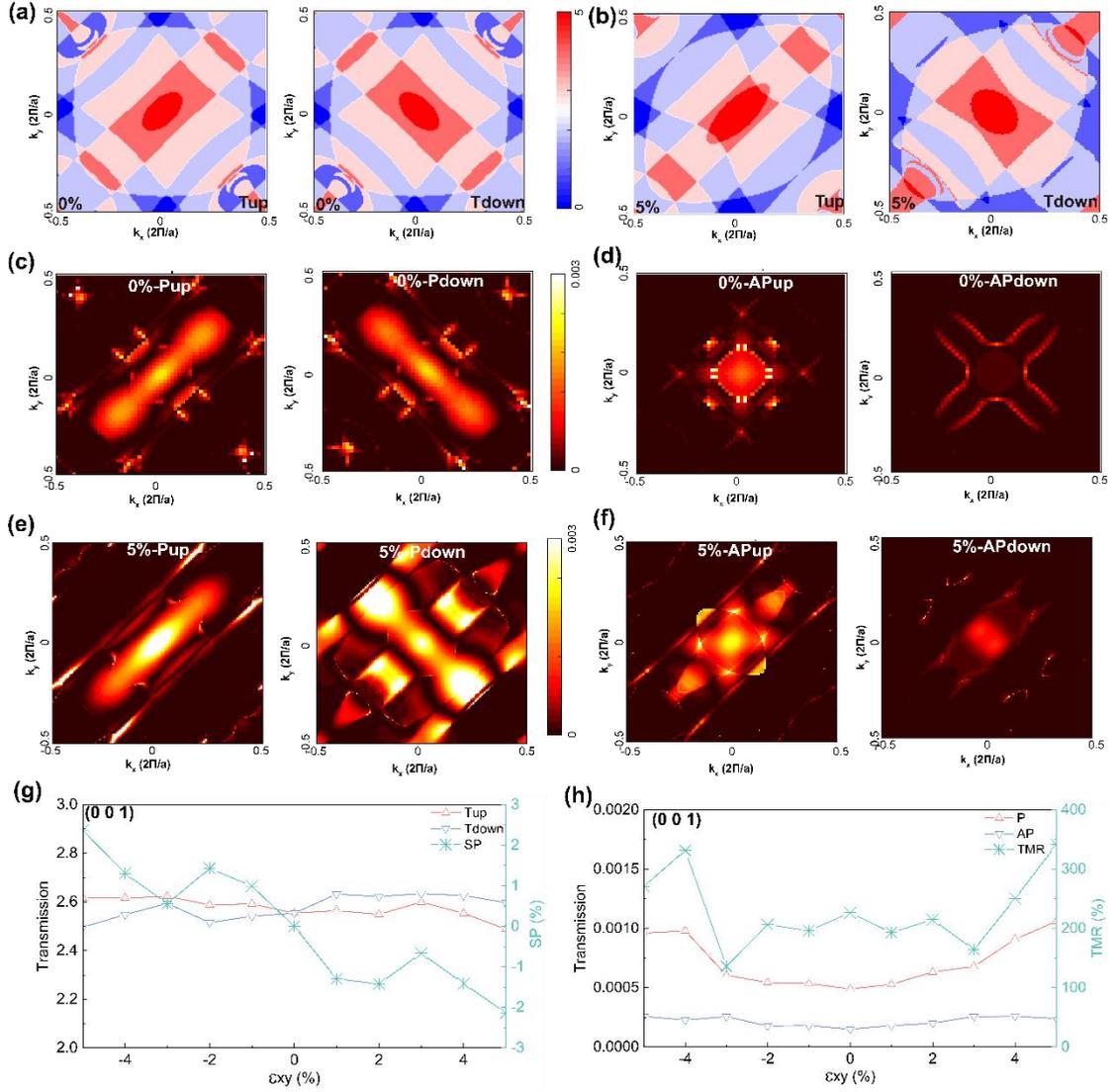

**Figure 3.** The number of $\vec{k}_\parallel$-resolved -resolved conduction channels in the 2D Brillouin zone of (a) intrinsic RuO$_2$ with (b) 5% shear strain for spin up $N_\parallel^\uparrow$ (left) and spin down $N_\parallel^\downarrow$ (left) (right) along [001] crystallographic directions. Spin $\vec{k}_\parallel$-resolved transmission in P and AP states of (c)-(d) intrinsic RuO$_2$ with (e)-(f) 5% shear strain for spin up (left) and spin down (right) along [001] crystallographic directions. (g) Variations in spin-resolved electron transport and spin polarization with shear strain magnitude and direction in bulk RuO$_2$ along [001] crystallographic directions. (h) Transmission in P and AP states and TMR as a function with shear strain magnitude and direction in RuO$_2$/TiO$_2$/RuO$_2$ MTJ along [001] crystallographic directions.

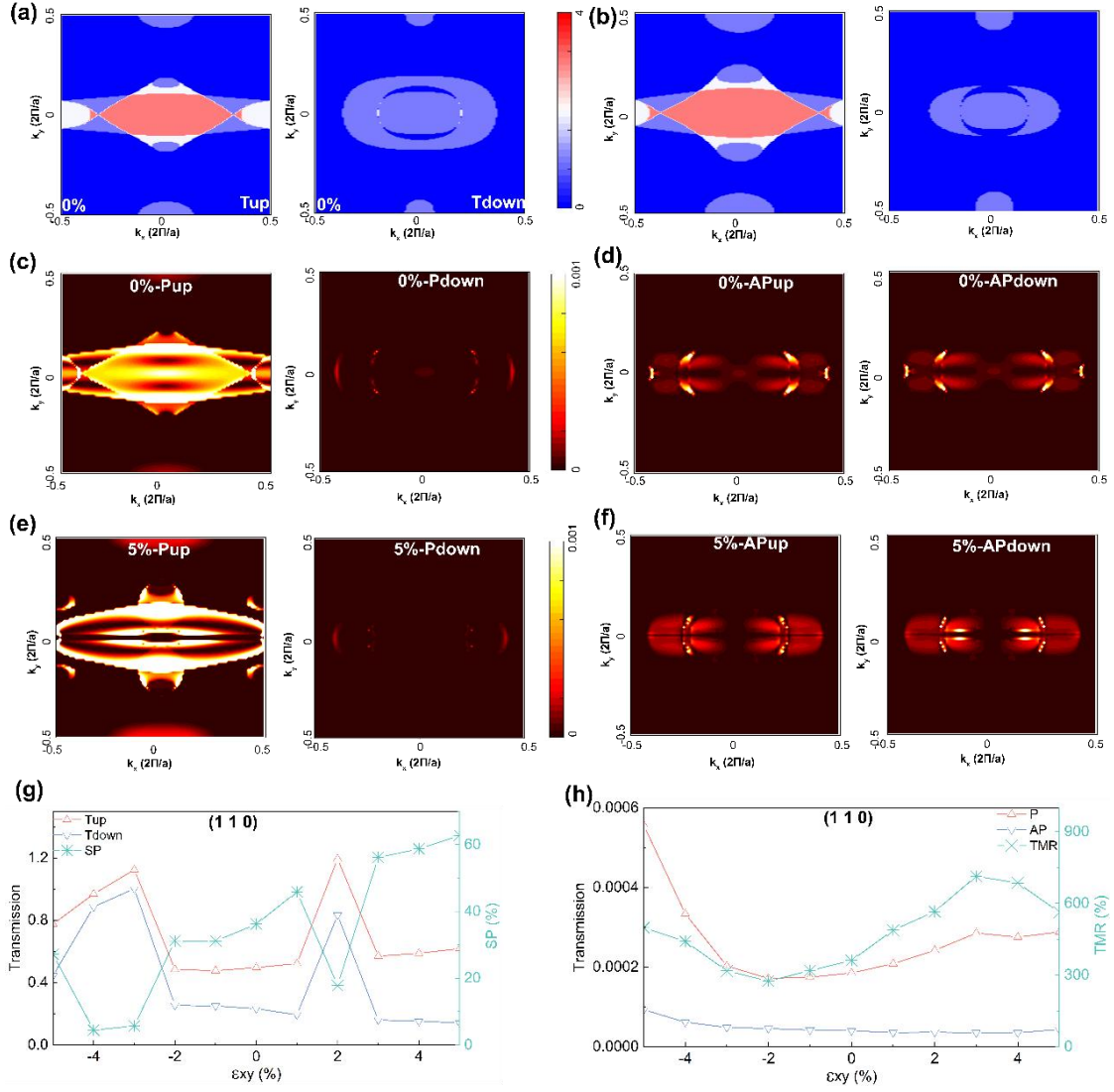

**Figure 4.** The number of $\vec{k}_\parallel$-resolved -resolved conduction channels in the 2D Brillouin zone of (a) intrinsic RuO$_2$ with (b) 5% shear strain for spin up $N_\parallel^\uparrow$ (left) and spin down $N_\parallel^\downarrow$ (left) (right) along [110] crystallographic directions. Spin $\vec{k}_\parallel$-resolved transmission in P and AP states of (c)-(d) intrinsic RuO$_2$ with (e)-(f) 5% shear strain for spin up (left) and spin down (right) along [110] crystallographic directions. (g) Variations in spin-resolved electron transport and spin polarization with shear strain magnitude and direction in bulk RuO$_2$ along [110] crystallographic directions. (h) Transmission in P and AP states and TMR as a function with shear strain magnitude and direction in RuO$_2$/TiO$_2$/RuO$_2$ MTJ along [110] crystallographic directions.

For the intrinsic RuO$_2$ [110] path, the electronic bands exhibit substantial spin splitting. Consequently, the RuO$_2$/TiO$_2$/RuO$_2$ (110) MTJ exhibits a very large TMR. Analysis of the intrinsic RuO$_2$ spin-resolved conduction channel count reveals that the number of spin-up channels is significantly higher than that of spin-down channels, resulting in a considerable spin polarization (36%). Upon the application of positive shear strain, the spin polarization increases markedly, peaking at 63%. Correspondingly, the TMR of the tunnel junction also rises significantly, from 361% to 713%. In sharp contrast, applying negative shear strain reduces the spin polarization, leading to a decrease in TMR to 275%. Simultaneously, we find that applying shear strain can also substantially enhance the conductance of the $RuO_2/TiO_2/RuO_2$ (110) MTJ.

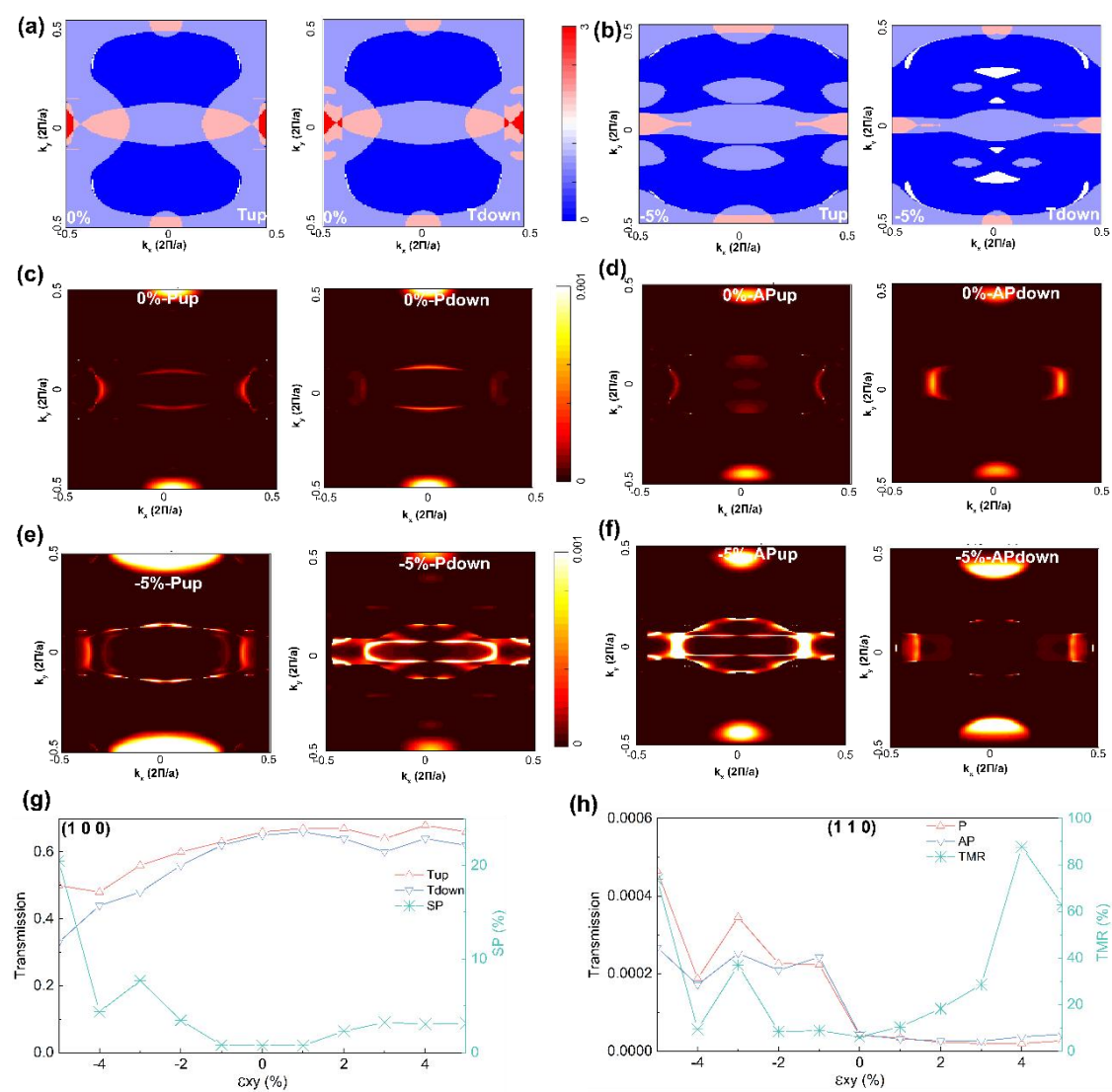

**Figure 5.** The number of $\vec{k}_\parallel$-resolved -resolved conduction channels in the 2D Brillouin zone of (a) intrinsic $RuO_2$ with (b) 5% shear strain for spin up $N_\parallel^\uparrow$ (left) and spin down $N_\parallel^\downarrow$ (left) (right) along [100] crystallographic directions. Spin $\vec{k}_\parallel$-resolved transmission in P and AP states of (c)-(d) intrinsic $RuO_2$

with (e)-(f) 5% shear strain for spin up (left) and spin down (right) along [100] crystallographic directions. (g) Variations in spin-resolved electron transport and spin polarization with shear strain magnitude and direction in bulk $RuO_2$ along [100] crystallographic directions. (h) Transmission in P and AP states and TMR as a function with shear strain magnitude and direction in $RuO_2/TiO_2/RuO_2$ MTJ along [100] crystallographic directions.

For other high-symmetry paths in intrinsic $RuO_2$, the band structure exhibits no spin splitting. Concurrently, the spin-resolved conduction channels in momentum space are completely degenerate and equal in number. Consequently, $RuO_2/TiO_2/RuO_2$ MTJ based on these paths fail to exhibit a magnetoresistance effect. This significantly complicates the experimental fabrication of such MTJs, requiring strict epitaxial growth along specific crystallographic orientations. Here, taking the $RuO_2$ (110) orientation as an example, the aforementioned band structures reveal pronounced spin splitting along the Γ-X path upon the application of shear strain. Furthermore, in momentum space, a marked disparity emerges in the number of conduction channels between spin-up and spin-down electrons as the strain increases, with the spin polarization being tunable up to 20%. Consequently, the $RuO_2/TiO_2/RuO_2$ (110) MTJ under shear strain exhibits significant TMR, reaching a maximum value of 88%.

**Discussion**

Note that the TMR of the $RuO_2/TiO_2/RuO_2$ MTJ across the different orientations described above does not exhibit a strict proportionality to the spin polarization of $RuO_2$. This discrepancy arises because shear strain may also alter the $RuO_2/TiO_2$ interface or the barrier width within the $TiO_2$ layer – factors that notably modulate the TMR. For the three orientations of $RuO_2/TiO_2/RuO_2$ MTJ we constructed, the transport properties were studied only for specific $TiO_2$ barrier thicknesses. However, moderately increasing the barrier width can significantly enhance the TMR of the MTJ, even by several orders of magnitude, as reported in our previous work. Furthermore, this $RuO_2$-based non-alter compensated magnet can also be utilized in other MTJ configurations: For instance, in $RuO_2$/barrier/FM structures, where $RuO_2$ serves as the pinned layer; selecting a FM with high spin polarization can substantially boost the TMR. Alternatively, it can be employed in AFM/barrier/$RuO_2$ junctions, where $RuO_2$ acts as the free layer and the AFM comprises altermagnets or non-collinear AFM with momentum-space spin splitting. It can even be integrated with ferroelectric materials in multiferroic tunnel junctions to enable ultralow-power spintronic devices. Additionally, our study reveals that this shear-strain strategy for achieving a non-alter compensated magnet is broadly applicable to other d-wave altermagnets, including insulating altermagnets like $MnF_2$. Looking ahead, we aim to explore more strategies for breaking the rotational symmetry of altermagnets to modulate the non-alter compensated magnet, and we anticipate extending this approach to other altermagnet types, such as g-wave and i-wave systems.

**Conclusion**

Overall, we have investigated the application potential of uncompensated magnets in AFMTJ by density functional theory (DFT) to applying shear strain to d-wave altermagnets, such as $RuO_2$ and $MnF_2$, modulates them into an uncompensated magnetic state, characterized by band structures exhibiting non-alter spin splitting across the entire momentum space—distinct from their pristine altermagnetic behavior. This phenomenon arises because shear strain breaks the rotational symmetry inherent to altermagnets. Such symmetry breaking consequently disrupts the magnetic moment compensation between opposite sublattices and induces a weak magnetic moment. Furthermore, the direction of the applied shear strain reverses the material's magnetization. To further explore the application potential of these uncompensated magnets in tunnel junctions, we designed $RuO_2/TiO_2/RuO_2$ MTJ with three crystallographic orientations (001, 110, and 100), using $RuO_2$ as a representative case. For the (001) and (110) orientations, shear strain enhances the spin polarization of $RuO_2$. Notably, for the $RuO_2$ (110) orientation, it significantly increases the spin polarization from 21% to 61%, thereby leading to a marked enhancement of the TMR in the corresponding MTJ under strain. Conversely, for orientations like (100) or other spin-degenerate paths (where intrinsic $RuO_2$-based MTJs exhibit no magnetoresistance effect), the MTJ exhibits appreciable TMR under applied shear strain. Our work aims to provide valuable insights for exploring magnetic electronic structures and developing novel magnetic devices.

## ACKNOWLEDGMENTS

We gratefully acknowledge the financial support received from the National Key Research and Development Program of China (Grant No. 2022YFA1602701), the National Natural Science Foundation of China (Grants No. 12227806, Grants No. 12204364 and Grant No. 12327806).

## Author contributions

The Fangqi Liu and Yanrong Song contributed equally to this work.

## Data availability

All relevant data are available from the authors.

## Notes

The authors declare no competing financial interest.